\documentstyle[11pt,newpasp,twoside]{article}
\def\emp)hasize#1{{\sl#1\/}}

\def\edcomment#1{\iffalse\marginpar{\raggedright\sl#1\/}\else\relax\fi}
\reversemarginpar

\begin{document}
\title{Possible re-acceleration regions above the inner gap
 and pulsar gamma-ray emission}
\author{B.H. Hong$^{1,2,3}$,
        G.J. Qiao$^{1,2}$, B. Zhang$^{1,2}$,
        J.L. Han$^{1,3}$,
        R.X. Xu$^{1,2}$}
\affil{$^1$Beijing Astrophysical Center, CAS-PKU, Beijing 100871, China\\
$^2$Astronomy Department, Peking University, Beijing 100871, China\\
$^3$National Astronomical Observatories, CAS, Beijing 100080, China}

\begin{abstract}
We propose that the violent breakdown of the Ruderman-Sutherland type
vacuum gap above the pulsar polar cap region may produce copious primary
particles and make the charge density above the gap in excess of the 
Goldreich-Julian density. Such an excess residue of charge density may
arise some re-acceleration regions above the gap due to the 
space-charge-limited flow mechanism. 
Such re-acceleration regions, if exist, may
accelerate particles to carry away more spin-down energy, and may have
some implications for pulsar $\gamma$-ray emission.

\end{abstract}

Much has been done to investigate the formation of pulsar inner accelerators
near the polar cap region (Ruderman \& Sutherland 1975 [RS75]; Arons \& 
Scharlemann 1979; Harding \& Muslimov 1998) via various charge 
deficit mechanisms. Here we propose another possible particle acceleration 
mechanism in the pulsar inner magnetospheres via a {\em charge excess} 
mechanism. The violent breakdown of the vacuum gap (RS75)
may produce extra charges to make the charge density above the gap deviated 
from the Goldreich-Julian (1969) density, so that the 
space-charge-limited flow mechanism may result in one or more 
re-acceleration regions above the gap.

\vspace{0.2cm}
{\bf Basic idea of the re-acceleration regions\ } 
Above the gap, the secondaries do not contribute to the pure charge 
density, so the net contribution of $\rho$ is just the primary charge 
density. In the previous polar cap acceleration models (RS75;
Arons \& Scharlemann 1979; Harding \& Muslimov 1998), it is assumed 
that $E_\parallel$ is completely screened.
However, a complete screening of the field 
requires both $E_\parallel=0$ and ${\rm d} E_\parallel/{\rm d} 
z=0$ to be achieved.
This is only {\em assumed} in the previous models.
In reality, both conditions can not be achieved simultaneously
at the very beginning, and there must be some oscillation and 
relaxation processes to readjust $\rho$ and $E_\parallel$ to achieve 
final screening. In principle, such a detailed process should be 
investigated explicitly. Here we present our first attempt to treat 
such a complicated process within the framework of the vacuum gap model 
(RS75). We argue that the charge density 
$\rho$ of the primary particles flowing out from the gap could be 
considerably in excess of $\rho_{\rm gj}$ if the breakdown is very violent,
e.g. $\rho=\xi\rho_{\rm gj}$ with $\xi>1$. 
In the near surface regime (RS75), Poisson equation 
could be approximately treated as one-dimensional, i.e. ${\rm d} E_\parallel/
{\rm d} z
=4\pi(\rho-\rho_{\rm gj})$. Within the gap, since $\rho=0$, and $\rho_{\rm 
gj}>0$,
the $E_\parallel$ gradient is negative, and
$E_\parallel\simeq (2\Omega B/c) (h-z)$ (RS75).
Above the gap, the possible excess of the charge density ($\rho>\rho_{\rm gj}$) 
actually changes the sign of the $E_\parallel$ gradient, so that
$E_\parallel$ starts to grow with the direction of $E_\parallel$ 
unchanged. Thus another {\rm re-acceleration} region is probably 
formed. Completely neglecting the screening fields of the pairs, our 
results show that a series of accelerators will be formed, each of 
which again limited by another pair formation front. At the top of
each re-acceleration region, there is also an additional charge 
excess
(denoted as $\xi_2$, $\xi_3$, etc.). The relation between the 
adjacent charge excess is approximately ${(\xi_3-1) / (\xi_2-1)}=
1+{h_2 / (R+h_1)}>1$, where $h_1$, $h_2$ are the heights of the first and 
second accelerators, and $R$ the stellar radius. We see that $\xi_3>\xi_2$,
which means that the charge deviations could not be converged.

A complete treatment of the re-acceleration region must include the
screening fields of the secondary pairs. Such fields are mainly determined 
by the number of the pairs, i.e., the multiplicity factor of the pairs with
respect to the primaries (defined as $\eta$).
It is possible that the re-acceleration fields could not be completely 
screened if $\xi$ is large enough, but $\eta$ is not large enough, and
the structure of the re-accelerator is decided by the competition of 
$\xi$ and $\eta$. As mentioned above, more pair formation fronts will 
be formed above each re-acceleration regions, so that the increase of 
$\eta$ is much faster than the increase of $\xi$. As a result, it is
unlikely to form too many re-acceleration regions. Very probably, only 
one or two re-acceleration regions can exist. Detailed quantitative 
study of such re-acceleration regions is in progress.

\vspace{0.2cm}
{\bf About the luminosity of gamma-ray emission \ }
Observationally, $\dot E_{\rm rot}/4\pi d^2$ is adopted as a parameter
to select $\gamma$-ray pulsar candidates (Thompson 1996), and among
the 8 known $\gamma$-ray pulsars, an empirical law of $L_\gamma \propto
(\dot E_{\rm rot})^{1/2}$ is found. 
Zhang \& Harding (1999) proposed a full polar cap cascade model
based on the acceleration model of Harding \& Muslimov (1998),
which can successfully interpret
both the $\gamma$-ray and X-ray luminosities of the spin-powered pulsars.
In principle, the full cascade picture can be also used in the vacuum
gap models. However, with the possible existence of multipole fields
near the surface (RS75), the energetics of 
the vacuum gap is usually not 
enough to interpret $L_\gamma$. 
If, however, one or more re-acceleration regions proposed here do exist 
above the gap, particles within these ``quasi-gaps'' can be also 
accelerated to highly relativistic energies to bring away more 
spin-down energy to $\gamma$-ray emission, which may interpret
the observed $L_\gamma$ of pulsars.

\end{document}